\documentclass{PoS}
\usepackage{enumerate}

\newcommand\lsim{\mathrel{\rlap{\lower4pt\hbox{\hskip1pt$\sim$}}
    \raise1pt\hbox{$<$}}}
\newcommand\gsim{\mathrel{\rlap{\lower4pt\hbox{\hskip1pt$\sim$}}
    \raise1pt\hbox{$>$}}}
\newcommand{\ba}{\begin{array}}
\newcommand{\ea}{\end{array}}
\newcommand{\nn}{\nonumber}

\newcommand{\be}{\begin{equation}}
\newcommand{\ee}{\end{equation}}
\newcommand{\bear}{\begin{eqnarray}}
\newcommand{\eear}{\end{eqnarray}}

\newcommand{\ket}{\,\rangle}
\newcommand{\bra}{\langle \,}

\newcommand{\cO}{{\cal O}}

\newcommand{\mF}{\mathcal{F}}

\newcommand{\mK}{\mathcal{K}}
\newcommand{\mL}{\mathcal{L}}
\newcommand{\mM}{\mathcal{M}}
\newcommand{\mO}{\mathcal{O}}

\newcommand{\Frac}[2]{\frac{\displaystyle #1}{\displaystyle #2}}
\newcommand{\Int}{\displaystyle{\int}}

\def\bat{\begin{array}{cc}}

\title{Resonances and loops: \\ scale interplay in the Higgs effective field theory }

\ShortTitle{R\&$\ell$: scale interplay in the HEFT}

\author{
\speaker{Juan Jos\'e Sanz-Cillero}
\thanks{This work was partly supported by the Spanish MINECO fund FPA2016-75654-C2-1-P.}
\\
Departamento de F\'\i sica Te\'orica and UPARCOS,
Universidad Complutense de Madrid,
\\
E-28040 Madrid, Spain
\\
E-mail:
\email{jjsanzcillero@ucm.es}}

\abstract{
Within the current situation where there seems to be a large mass gap between the Standard Model (SM) and new physics (NP) particles, the effective field theory (EFT) framework emerges as the natural approach for the analysis and interpretation of collider data. However, this large gap and the fact that (so far) all the measured interactions look pretty much SM-like does not imply that the linear Higgs representation as a complex doublet $\phi$ in the Standard Model Effective Field Theory (SMEFT) is always appropriate. Although there is a wide class of SM extensions (e.g., supersymmetry) that accept this linear description, this realization does not always provide a good perturbative expansion. The non-linear Higgs Effetive Field Theory (HEFT) (also denoted as Electroweak Chiral Lagrangian, EW$\chi$L) considers an EFT with a singlet scalar Higgs $h$ and the triplet of electroweak Goldstones $\omega^a$, and its organization according to a chiral expansion cures these issues (even if the analysis sometimes becomes slightly more cumbersome). Path integral functional methods allow one to compute the corrections to the next-to-leading order (NLO), $\cO(p^4)$, effective action: at tree level, the heavy resonances only contribute to the $\cO(p^4)$ low-energy couplings (or higher) according to a pattern that depends on their quantum numbers and is suppressed by their masses; at the one-loop level, all the new-physics corrections go to the  $\cO(p^4)$  effective action (or higher) and are suppressed by an intrinsic scale of the leading order (LO), $\cO(p^2)$, HEFT Lagrangian, which has been related in recent times with the curvature of the manifold of scalar fields (being the SM its flat limit). Thus, the contribution of the leading HEFT Lagrangian to the effective action might be understood as an expansion in its curvature. From the theoretical point of view, this two suppression scales (resonance masses and non-linearity/curvature) compete at NLO and may, in principle, be very different/similar depending on the particular beyond-SM (BSM) scenario and the observable at hand. This is what happens in Quantum Chromodynamics, where there are multiple scales and the relevant meson mass and the loop suppression scale factor depend on the channel at hand --and even have a different scaling with the number of colours--. Likewise, experiments do not tell us yet whether $\cO(p^4)$ loops are essentially negligible or not, as it is shown with a couple of examples. In summary, the HEFT extends the range of applicability of the SMEFT, which, although justified for a pretty wide class of BSM scenarios, introduces a bias in the data analyses and does not provide the most general EFT framework.
}

\FullConference{EPS-HEP 2017, European Physical Society conference on High Energy Physics, \\
		5-12 July 2017, 
		Venice, Italy}

\begin{document}

\section{A tale of two realizations}

In the case that a large mass gap with the heavier NP state $R$
is confirmed, as experiments seem to be indicating, the EFT approach
appears as the most convenient one for future data analysis and interpretation.
The aim of this talk is to discuss two questions that are often posed
in relation with the so-called non-linear HEFT (also denoted sometimes as EW$\chi$L):
\begin{itemize}

\vspace*{-0.cm}
\item{\it ``Since, experimetally, we deem to be so close to the SM,
is it not enough to consider the SMEFT framework
to parametrize NP effects at low energies?    }

\vspace*{-0.cm}
\item{\it ``Are not BSM loops essentially negligible?''}

\end{itemize}
The answer in both cases is {\it no}: the SMEFT is appropriate and loops are sub-subdominant only
in particular cases (although this includes a broad set of NP scenarios, such as supersymmetry),
as we will discuss.  
The two possible representations of the  
EFT based on the SM symmetries
for any BSM extension are:
\begin{itemize}

\item{\bf Linear (L) -- SMEFT:} the EFT is build in terms of the complex doublet with the general form
$\phi =\left(\begin{array}{c} \varphi^+\\ \varphi^0\end{array}\right)
= (1+H/v) U(\omega^a) \bra 0| \phi |0 \ket$
(with the vacuum expectation value $v=0.246$~TeV
and the unitary matrix $U$ containing the  electroweak (EW) Goldstones $\omega^a$),
leading to an EFT Lagrangian organized according to the
canonical dimension $D$ of the operators~\cite{deFlorian:2016spz,Brivio:2017vri}
\bear
\mL^{\rm (L)} &=& \mL_{SM\, (D\leq 4)} + \sum_{D>4} \mL_D = |D_\mu \phi|^2  +  \Frac{c_H}{\Lambda^2}
\left(\partial_\mu (\phi^\dagger \phi)\right)^2 + ...
\nn\\
&& = \Frac{ (v+H)^2}{4} \bra D_\mu U^\dagger D_\mu U\ket
+  \Frac{1}{2} (1 + P(H))\, (\partial_\mu h)^2
+ ...
\label{eq:L-Lag}
\eear

\item{\bf Non-linear (NL) -- HEFT:} the EFT is built in terms of 1 singlet $h$ and 3 EW Goldstones
$\omega^a$ encoded
in the NL representation provided by the unitary matrix $U(\omega^a)$,
leading to an EFT organized according the chiral dimension $d$ of the operators
($\mL_{SM}\subset \mL_{\cO(p^2)}$)~\cite{deFlorian:2016spz,R-LECs,HEFT-Lag}
\bear
&& 
\mL^{\rm (NL)} = \mL_{\cO(p^2)} + \mL_{\cO(p^{\geq 4})}\, ,
\label{eq:NL-Lag}
\\
&&\mL_{\cO(p^2)}
= \Frac{v^2}{4} \mF_C(h) \bra D_\mu U^\dagger D^\mu U\ket + \Frac{1}{2} (\partial_\mu h)^2 + ...\, ,
\qquad \mL_{\cO(p^4)} = \sum_k \mF_k(h) \mO_k\, ,
\nn
\eear
with $\mF_C(h)=1 + 2 a h/v + b h^2/v^2 +\cO(h^3)$ having an analytical expansion
around $h=0$.~\footnote{
The $a$ and $b$ terms provide the $hWW$ and $hhWW$ interactions, respectively,
with normalization $a^{SM}=b^{SM}=1$.}
\end{itemize}
It is possible to rewrite the SMEFT~(\ref{eq:L-Lag})
in the NL form~(\ref{eq:NL-Lag}):
\bear
&&\qquad {\rm \bf [L\to NL]:}\qquad \Frac{dh}{dH} = \sqrt{1+P(H)}
\quad \Longrightarrow \quad
h = \Int_0^H \sqrt{1+P(H)} \, dH\, .
\eear
Likewise, provided there exists an $SU(2)_L\times SU(2)_R $ fixed point $h^*$
with $\mF_C(h^*)=0$~\cite{R-HEFT}, it is always possible to revert
the complete HEFT Lagrangian~(\ref{eq:NL-Lag}) into the L form in terms of $\phi$
in~(\ref{eq:L-Lag}):
\bear
&& \hspace*{-2.7cm} {\rm \bf [NL\to L]:}\qquad \phi^\dagger \phi = (v+H)^2/2 = \Frac{v^2}{2} \mF_C(h) \, .
\eear

The issue is not whether we write our all-order EFT Lagrangian
in the form~(\ref{eq:L-Lag}) or~(\ref{eq:NL-Lag}),
the problem is to make sense of the EFT perturbative expansion in that realization~\cite{Brivio:2017vri}.
Thus, if one assumes the SMEFT~(\ref{eq:L-Lag}) and rewrites it in the $\mL^{\rm (NL)}$ form,
the deviations from the SM in the $hWW$ and $hhWW$ vertices,
respectively $\Delta(a^2)=a^2-1$ and $\Delta b = b-1$, have the precise form~\footnote{
There is another dimension-6 operator in the SMEFT, $c_T$~\cite{deFlorian:2016spz,Brivio:2017vri},
that modifies the Goldstone kinetic term once $\mL^{\rm (L)}$ is rewritten in the
$\mL^{\rm (NL)}$  form.
However, it does not contributes to $\mF_C(h)$ in $\mL_{\cO(p^2)}$,
but to the $a_0\mO_0$ Longhitano operator~\cite{HEFT-Lag},
NLO due to its large experimental suppression
in the $T$ oblique parameter~\cite{deFlorian:2016spz,R-LECs,oblique-res}.
}
\bear
&&\hspace*{-1.15cm} {\rm\bf [SMEFT\to HEFT]:}
\,\,\,  
\Delta(a^2) = - 2 c_H v^2/\Lambda^2 + ...  
,
\,
\Delta b = - 4 c_H v^2/\Lambda^2 + ...  
\Longrightarrow
2 \Delta(a^2) = \Delta b + ... 
\label{eq:db2da2}
\eear
where the dots stand for the  $\cO\left((v/\Lambda)^{n\geq 4}\right)$ contributions from
SMEFT operators with $D\geq 8$.
The next couple of examples shows the potential issues of the L representation:
\begin{enumerate}[i.)]
\item{\bf Dilaton Higgs models~\cite{Goldberger:2008zz}:} 
formulated in the NL representation~(\ref{eq:NL-Lag}),
they obey the constraint $\Delta (a^2) = \Delta b$.
Thus, there are large corrections in~(\ref{eq:db2da2})
from operators of dimension $D>6$
if one writes $\mL^{\rm (NL)}$ in the L form~(\ref{eq:L-Lag}),
leading to a breakdown of the SMEFT expansion~\cite{Brivio:2017vri}.

\item{\bf $SO(N)/SO(N-1)$ composite Higgs models (CHM)~\cite{MCHM}:}
Interestingly enough, the strongly interacting CHM
can always be rewritten in the SMEFT form~\cite{R-HEFT}.~\footnote{
In particular, these $SO(N)/SO(N-1)$ CHM obey the relation in Eq.~(\ref{eq:db2da2}). } 
However, its SMEFT $v/f$ expansion is poorly convergent
for $v\sim f$ at any energy $E$ (or not at all),~\footnote{
Note that the 95\%CL determination $a> 0.8$~\cite{Buchalla:2015qju}
implies a rather loose lower
bound for the $SO(N)/SO(N-1)$ spontaneous symmetry breaking scale~\cite{MCHM}
$f=v/\sqrt{1-a^2}>0.4$~TeV for the CHM scale.  }
while its $\mL^{\rm (NL)}$ gives an EFT with loops suppressed by powers
of $E^2/(16\pi^2 v^2)$~\cite{deFlorian:2016spz},  
at least~\footnote{
Explicit computations show that even if individual loop diagrams are suppressed in this way,
the full one-loop amplitude shows a much stronger suppression
$(1-a^2)E^2/(16\pi^2 v^2)=E^2/(16\pi^2 f^2)$ for the HEFT derived from the $SO(5)/SO(4)$ CHM~\cite{Delgado:2014jda}.
}  valid for $E\ll 4 \pi v$.

\end{enumerate} 

\section{HEFT chiral expansion and the competition between scales: loops vs. tree-level}

$\mL^{\rm (NL)}$  
is sorted out according
to the chiral dimension of its operators, based on the scaling
$D_\mu, \, m_{\rm Bos} , \, m_{\rm Fer} ,\, y_{\rm Fer}, g^{(')}
\sim \cO(p)$~\cite{deFlorian:2016spz},
such that the observables have an expansion $\mM\sim  \sum_n \mM_n (p^2)^n$,
where $p$ stands for the appropriate combination of soft scales of the EFT ($k_i^\mu,\, m_j$).
For instance, 
\bear
\mM_{2\to 2} \,\,\, &=&\,\,\,  \underbrace{ \Frac{ a_{_\mM}  p^2}{v^2} }_{\rm LO \, (tree)}
\quad + \quad
\underbrace{  \Frac{\mF_{_\mM}  \, p^4}{v^4}   }_{\rm NLO \, (tree)}
+ \underbrace{ \left( \Frac{ p^4 \Gamma_{_\mM} }{16\pi^2 v^4}   \ln\Frac{\mu}{p}
+ ...\right)}_{\rm NLO\, (1-loop)}
\quad + \quad \underbrace{\cO(p^6)}_{\rm NNLO + ... }\, ,
\label{eq:amp-chiral-exp}
\eear
where $a_{_\mM}$ ($\cO(p^2)$) and $\mF_{_\mM}$  ($\cO(p^4)$) are
the corresponding combination of renormalized couplings, with $\Gamma_{_\mM}$
providing their running~\cite{Guo:2015isa,R-HEFT}. 
Actually, in a rather baroque way, the SM is also organized in a chiral expansion,
with
$\frac{p^2}{16\pi^2 v^2} = \frac{g^{(')\, 2}}{16\pi^2},\,
\frac{\lambda_{\rm Fer}^2}{16\pi^2},\, \frac{\lambda}{16\pi^2}$.~\footnote{
At LO $g^2= 4 m_W^2/v^2$, $g^{'\, 2}= 4 (m_Z^2-m_W^2)/v^2$, 
$\lambda =  m_H^2/(2v^2)$ and
$y_{\rm Fer}^2=m_{Fer}^2/(2v^2)$. }
Since the latter  $p$'s are not external
momenta but coupling constants, the SM amplitudes (arranged like Eq.~(\ref{eq:amp-chiral-exp}))
do not diverge like a power of the energy. 
Obviously, both in SM and BSM scenarios, particular kinematic regimes 
may require further refinements of the EFT, as there may be further hierarchies
between EFT scales.

Path integral functional methods allow one to extract this two kinds of contributions: 
\begin{itemize}

\item{\bf NLO tree-level:} In order to integrate out the heavy fields $R$~\cite{R-LECs},
1) one extracts their equation-of-motion solutions (EOMS) $R_{\rm cl}$
for the underlying high-energy Lagrangian $\mL_{\mbox{\tiny High-E}}[R, \mbox{\tiny light}]$,
2) substitutes them therein,  
and 3) considers its low-energy expansion $R_{\rm cl} \sim
\frac{G_R}{M_R^2}  \, \chi_R[\mbox{\tiny light}]$: 
\bear
&&\hspace*{-2.2cm}
\mL^{\rm (NL),\, tree}_{NLO} = \mL_{\mbox{\tiny High-E}}\big[
\, R_{\rm cl}[\mbox{\tiny  light}],\, \mbox{\tiny light}\, \big] 
\stackrel{p\ll M_R}{\sim}  \Frac{G_R^2}{M_R^2}  \big( \chi_R[\mbox{\tiny  light}] \big)^2 + ...
\, =  \, \sum_k \mF_k \mO_k + ... 
\, , \,\,\, \mbox{with } \mF_k\approx \Frac{G_R^2}{M_R^2}
\, ,
\eear
where the dots stand for $\cO(p^{d\geq 6})$ terms.
Since chiral symmetry 
demands $\chi_R[\mbox{\tiny light}] \sim p^2$~\cite{R-LECs},
its contributions only go to $\mL_{ \cO(p^{d\geq 4})}$ after integrating out each $R$, 
preventing large deviations from the SM in $\mL_{\cO(p^2)}$
and leading to a particular pattern of $\cO(p^4)$ HEFT couplings~\cite{R-LECs}. 

\item{\bf NLO 1-loop:} Considering fluctuation $\eta_i$ of the HEFT fields around their EOMS,
and expanding $\mL_{\cO(p^2)}$  in powers of $\eta_i$, 
the $\cO(\eta^2)$ terms yield the one-loop contributions~\cite{R-HEFT,Guo:2015isa}~\footnote{
This refers to purely bosonic one-loop contributions.
In order to compute the fermion one-loop contribution one needs
to consider also their fluctuation around EOMS and generalize the trace $tr$
to a supertrace $str$~\cite{fermion-loops}.
The computations are performed in dimensional regularization
with space-time dimension $d\to 4$ and renormalization scale $\mu$.   } 
\bear 
&& 
\hspace*{-3.5cm}
\Int{\rm d^dx}\,\, \mL^{\rm (NL),\, 1\ell}_{NLO} = \Frac{i}{2}{\rm tr}\log{  \left(
-\Frac{\delta^2 \mL_{\cO(p^2)}}{\delta\eta_i\, \delta\eta_j}
\right)} \,=\,
\, - \Frac{ \mu^{d-4} }{16\pi^2 (d-4)} \Int{\rm d^dx} \sum_k \Gamma_k \mO_k {\rm \,+\,  finite}\, ,
\label{eq:1-loop}
\eear
These one-loop contributions are $\cO(p^4)$~\cite{R-HEFT,Guo:2015isa,fermion-loops}
and the BSM corrections are suppressed by $\Delta\mK^2=\left((\mF_C')^2 /\mF_C-4\right)$
and $\Omega=\left( 2 \mF_C''/\mF_C - (\mF_C'/\mF_C)^2\right)$~\cite{Guo:2015isa}, vanishing
in the SM limit. It is interesting to remark the geometrical analysis in~\cite{R-HEFT},
where the  manifold of scalar fields $(h,\vec{\omega})$
has an associated metric $g_{ij}   
=$diag$\left\{\mF_C(h) g_{ab}(\omega) , 1 \right\}$ in the HEFT.~\footnote{
The indices $i,j$ run through the four scalars while
$a,b$ are restricted to the $\vec{\omega}$ coordinates.
The $SO(4)$ non-linear sigma model metric $g_{ab}(\vec{\omega})$ is provided, e.g., in~\cite{R-HEFT}.
Different coordinate choices give different metrics: e.g., in the SMEFT,
$g_{ij}=$diag$\left\{(1+H/v)^2 g_{ab}(\vec{\omega}) , 1+P(H) \right\}$.
}
Interestingly, the corresponding curvature~\footnote{
All the discussion in these proceedings refers to the
associated Riemann $ \mathbb{R}_{ijk\ell}$, Ricci $\mathbb{R}_{ij}$ and scalar curvature
$\mathbb{R}$ tensors~\cite{R-HEFT}, which will loosely denote as ``curvature'' $\mathbb{R}$.  }
$\mathbb{R}$ is proportional
to $\Delta \mK^2$ and $\Omega$.   
Thus, the scale suppressing the one-loop corrections
with respect to the LO is $\Lambda^{-2} = \mathbb{R}/(4\pi)^2$. 
Barring non-derivative operators,
the theory becomes non-interacting in the flat limit (the SM) --both at the tree and loop level--.
This points out that the true expansion scale of the EFT
in general is not the one that explicitly appears suppressing
the EFT operators and individual diagrams (e.g., $4\pi v$ in the HEFT),
which depends on the particular choice of coordinates for the scalar fields;
the low-energy amplitudes derived from $\mL_{\cO(p^2)}$
rather seem to be organized via an expansion in
a more intrinsic and coordinate-independent scale, $\mathbb{R}$:    
\bear 
\mbox{\bf [HEFT$\big|_{\mL_{\cO(p^2)}}$]:}
\qquad\qquad
\mM \,\,\, &=& \mathbb{R} p^2\,\,\,  \,+\, \,\,\, \Frac{ \mathbb{R}^2 p^4}{(4\pi)^2} \,\,\, \, +\,\,\, \,
\cO(p^6)\, .
\eear 

\end{itemize}

\section{Phenomenological examples of scale interplay}

Depending on its quantum numbers,  
the observables are contributed by resonances 
with different characteristics (lighter, heavier, narrow, broad or even absence
of resonant contributions) which may enter in competition with NLO loops of a similar size,
as in happens in Quantum Chromodynamics. 
This is specially relevant
in quantities where the $\cO(p^2)$ is absent and start at NLO~\cite{deFlorian:2016spz}: 
\begin{itemize}

\item{\bf One-loop resonance computation of
the oblique $S$ and $T$ parameters~\cite{oblique-res}: }
the two-Weinberg sum-rule scenario in \cite{oblique-res}
yielded an $\cO(p^4)$ tree-level contribution suppressed
at the 95\% CL by a scale $\Lambda\gsim 5$~TeV
(given by the vector and axial-vector resonance
mass contributions $-(M_V^2 +M_A^2)/(4M_V^{2}M_A^{2})$  to the HEFT coupling
$a_1$~\cite{deFlorian:2016spz}) and the $\cO(p^4)$ loops suppressed by
$\Lambda\gsim 30$~TeV (related to the chiral log coefficient $(a^2-1)/(192\pi^2 v^2)$~\cite{oblique-res,Delgado:2014jda}).

\item{\bf $\gamma\gamma\to ZZ$~\cite{Delgado:2014jda}:}
Based on the Run-1  
fit~\cite{Buchalla:2015qju},
the $\cO(p^4)$ tree-level contribution is suppressed
at the 95\% CL by a scale $\Lambda\gsim 1.4$~TeV
(provided by the combination $2 a c_{\gamma}/v^2$~\cite{Delgado:2014jda})
and the $\cO(p^4)$ loops suppressed by
$\Lambda\gsim 2.6$~TeV (given by $(a^2-1)/(4\pi^2 v^2)$~\cite{Delgado:2014jda}).

\end{itemize}
In summary: BSM extensions may, in general, contain more than one hard scale.
Which one is the lightest one in the low-energy EFT depends on the particular case.
The HEFT organized through a chiral expansion provides a systematic approach to tackle these issues,
valid up to $E= 4 \pi v$ (or higher energies~\footnote{
Due to subtle cancelations in close-to-SM scenarios 
the HEFT might be valid up to energies way higher than $4\pi v$
as it happens, e.g., in the SM or the $SO(5)/SO(4)$ CHM~\cite{MCHM}. In the latter,
individual 1-loop diagrams in the HEFT representation have
a suppression $\cO(p^2/(16\pi^2 v^2)$ but
a strong cancellation shows up after summing them up,
recovering the  $E^2/(16\pi^2 f^2)$ loop suppression one obtains in the underlying
$SO(5)$ non-linear sigma model diagram by diagram~\cite{Delgado:2014jda}. 
}) and is expected to lead to promising NP collider signals
in future years~\cite{collider-signals}.

\end{document}